\documentclass[12pt,preprint]{aastex}

\newcommand{\eg}{{\it e.g.,}}

\begin{document}

\title{The Host Galaxy and the Environment of the z = 1.195 Quasar 
3C\,190\footnotemark[1]}

\footnotetext[1]{Based on observations made with the NASA/ESA Hubble
Space Telescope, obtained from the data archive at the Space Telescope
Science Institute, which is operated by the Association of Universities
for Research in Astronomy, Inc., under NASA contract NAS 5-26555.}

\author{Alan Stockton\altaffilmark{2,3}}
\affil{Institute for Astronomy, University of Hawaii, 2680 Woodlawn
 Drive, Honolulu, HI 96822}

\author{Susan E. Ridgway\altaffilmark{2,3}}
\affil{Department of Physics and Astronomy, Johns Hopkins University,
Homewood Campus, Baltimore, MD 21218}

\altaffiltext{2}{Visiting Astronomer, W.M. Keck Observatory, jointly operated
by the California Institute of Technology and the University of California.}

\altaffiltext{3}{Visiting Astronomer, Canada-France-Hawaii Telescope, operated 
by the National Research Council of Canada, the Centre National de la Recherche 
Scientifique de France and the University of Hawaii.} 

\begin{abstract}
We present deep imaging and spectroscopic observations of the $z=1.195$
quasar 3C\,190 and its immediate environment.  The quasar is surrounded by an 
asymmetric low-surface-brightness
envelope in which there are also several galaxies with characteristic
dimensions of 3--5 kpc.  Some of these form a linear structure some
50 kpc long in projection, offset from the quasar and almost perpendicular
to the radio axis.  Spectroscopy of this feature indicates the presence
of a starburst component as well as stars a few $\times10^8$ years old.
Velocities of the emission-line gas associated with \ion{H}{2} regions in
the linear feature are difficult to reconcile with any reasonable model 
that is also consistent with the observed morphology.  One possible
scenario is an edge-on disk in an extremely massive ($\sim10^{12} M_{\sun}$) 
halo; another is two independent chain galaxies, with the apparent alignment
being fortuitous.  Taking into account all of the types of extended material we
find in the vicinity of 3C\,190, we appear to be witnessing
a relatively brief phase in the development of a spheroidal galaxy, in which
merging is proceeding nearly simultaneously in a variety of different regimes.
This system may be one of the clearest examples yet found for the mechanism by
which many elliptical galaxies and bulges of early-type spirals form in
the early Universe.
\end{abstract}

\keywords{galaxies: interactions---galaxies: evolution---quasars: 
individual (3C\,190)}

\section{Introduction}

Strong radio sources at low redshift are invariably found in massive
galaxies with dominant spheroidal components.  At higher redshifts
($z\gtrsim1$), the observed optical/IR morphologies appear to be less regular:  
they often show elongations in the direction aligned with the radio
axis \citep[the ``alignment effect'':][]{cha87,mcc87,cha90}, and they tend 
also to be complex, comprising several discrete components.  Nevertheless, 
there is little doubt that powerful radio galaxies and quasars at high 
redshifts are among the precursors
of the elliptical-galaxy population at later epochs; in particular, the
tight correlation between black hole mass and stellar velocity dispersion
in bulge populations \citep{fer00,geb00} points strongly in this direction.
Provided that the aligned structure and the multiple components are not 
simply effervescent froth in comparison to the galaxy as a whole, they 
may tell us something about the formation and evolution of elliptical galaxies.

The $z=1.195$ quasar 3C\,190 has been classified as a compact-steep-spectrum
(CSS) radio source; it is also a member of the class of ``red'' quasars,
whose optical/IR continua fall steeply towards shorter wavelengths
\citep{smi80,sim00}.
These two properties have each often been taken to indicate systems
having high densities of gas and dust.  We discuss here imaging and 
spectroscopic observations of 3C\,190 that show that it lies in an unusually 
rich environment and that we may be witnessing one particular example of
an important phase in the process of formation of massive spheroids.

\section{Observations and data reduction}
\subsection{HST Imaging}
The {\it HST} images were obtained with WFPC2, with the F702W filter.  
In the rest frame of the quasar, this filter covers the region from
2700 \AA\ to 3580 \AA\ (half-transmission points), which will generally
be free of strong emission, except for broad \ion{Mg}{2} emission in
the quasar itself.  The system response is about 20\% of the peak at
the wavelength of the [\ion{O}{2}] $\lambda3727$ doublet.
The quasar was placed close to the center of WFC3, and 8 1000 s exposures
were obtained in a 4-point dither pattern, moving the telescope after
every pair of exposures.  After standard pipeline reduction, the individual
exposures were processed as described by \citet{rid97}; final
averaging of the images was done with the STSDAS task {\it gcombine}.

\subsection{Groundbased Imaging}
We have obtained imaging of 3C\,190 with the Keck I telescope and
the Near-Infrared Camera \citep[NIRC;][]{mat94}, with the Keck II telescope
and the Low-Resolution Imaging Spectrometer \citep[LRIS;][]{oke95}, and
with the Canada-France-Hawaii telescope (CFHT) and the Subarcsecond Imaging
Spectrograph (SIS).

The NIRC imaging, with the $K'$ filter, consisted of a total of 1080 s of 
exposure on 3C\,190 and 72 s of exposure on a brighter PSF reference star.
The observations were processed using standard IR reduction techniques.  The
flat-fielding was done using twilight-sky images, and the calibration
was determined from observations of standard stars GD\,71 and M\,16-A14 
\citep{cas92}.

The LRIS imaging observations used a Schott RG-850 filter, which, together
with the Tektronix CCD cutoff, gave a bandpass ($\lambda_c=9010$ \AA, 
FWHM = 1000 \AA) very similar to that of
the Sloan $z'$ band.  A total of 4200 s of useful exposure was obtained.
The CCD images were reduced via standard IRAF tasks, using a sky flat
derived from a median average of the dithered exposures, with objects
masked out.  The calibration was determined from observations of the
spectrophotometric standard star Feige\,67 \citep{mas90}. The weighted
mean flux density of the standard was found by integrating the flux density 
of the standard over the normalized filter profile. 

The SIS imaging on CFHT used a filter with a FWHM of 35 \AA\ centered at
8196 \AA\ to isolate the [\ion{O}{2}] $\lambda3727$ emission doublet at
the redshift of 3C\,190.  The total exposure was 9300 s.  Again, standard
CCD reduction procedures were used, with a flat-field determined from
exposures on a tungsten-halogen source illuminating the inside of the dome 
and a calbration from images of the spectrophotometric standard star 
G191B2B \citep{mas90}.  While we have not attempted to subtract off the 
continuum component, the narrow bandpass ensures that there is very little
continuum contamination.

\subsection{Keck LRIS Spectroscopy}
On four occasions, we obtained spectroscopy either of 3C\,190 or of material 
surrounding it with LRIS on Keck II.  The detector was a Tektronix 
$2048\times2048$ CCD.
All spectroscopy was done using 1\arcsec-wide slits; the remaining
details are given in Table~\ref{specparam}.  
\begin{center}
\begin{deluxetable}{lclcccc}
\tablecaption{Spectroscopic Observations \label{specparam}}
\tablehead{\colhead{Object} & \colhead{PA} & \colhead{Offset}
& \colhead{Dispersion} & \colhead{Integration} & \colhead{$\lambda_{cen}$} & 
\colhead{UT Date} \\
\colhead{} & \colhead{(deg)} & \colhead{(arcsec)}
& \colhead{(\AA\ pixel$^{-1}$)} & \colhead{(s)} & \colhead{(\AA)} & 
\colhead{} }
\startdata
Linear Feature &  $-23.5$   & \phn1.6 W  & 1.28 & 3600 & 8012 & 97 Oct 06\\
Galaxies to N & $-88.2$ & \phn2.2 N  & 1.28 & 7200 & 7995 & 98 Feb 15\\
3C\,190 &  $-88.2$   & \phn0.0   & 1.28 & \phn600 & 7995 & 98 Feb 15\\
Linear Feature & $-23.5$ & \phn1.6 W & 1.28 & 6000 & 7999 & 98 Feb 17\\
Linear Feature & $-23.5$ & \phn1.6 W & 2.44 & 4800 & 6351 & 98 Mar 21\\
3C\,190 & $-23.5$ & \phn0.0 & 2.44 & \phn173 & 6351 & 98 Mar 21\\
3C\,190 & \phs\phn0.0  & \phn0.0 & 2.44 & 1200 & 6350 & 98 Apr 06\\
\enddata
\end{deluxetable}
\end{center}

\section{Results and Interpretation}
\subsection{Imaging}
\begin{figure}[p]
\plotone{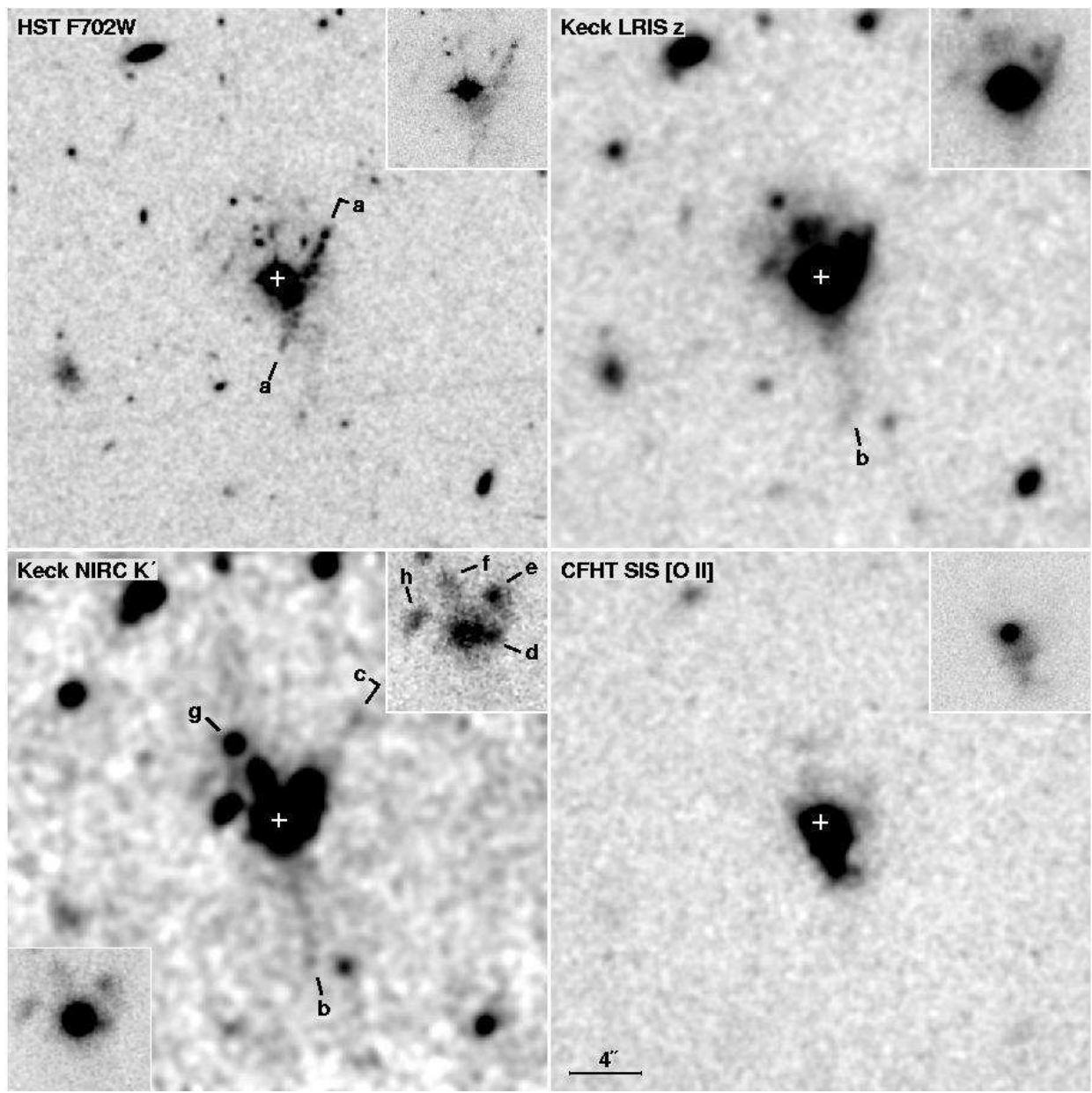}
\caption{Images of the field of 3C\,190.  The labels indicate the telescope,
instrument, and filter.  The main panels have been smoothed slightly to
show low-surface-brightness features better; insets in the upper-right
corners show unsmoothed versions at lower contrast.  The NIRC $K'$ image
has had the quasar PSF removed in the main panel and upper-right inset.  
The lower-left inset shows the field before PSF removal. The white crosses
indicate the quasar position.}\label{3c190mos} 
\end{figure}

The results of both the {\it HST} and the ground-based imaging are shown in
Fig.~\ref{3c190mos}.  The {\it HST} WFPC2 image and the Keck NIRC $K'$
image have been described briefly by \citet{sto99}.
The most striking feature in the {\it HST} WFPC2 image is the linear
alignment of luminous knots ($a$), extending south-southeast to north-northwest.
This feature does not pass through the quasar image, but its centerline comes
within $\sim1\farcs5$ at its closest point on the southwest side of the
quasar.  The three ``knots'' apparent on the northwest end of the linear
feature each have $m_{AB 6900}\approx24$, corresponding to 
$M_U\approx-21$ (we assume $H_0=75$ km s$^{-1}$ Mpc$^{-1}$, $\Omega_m=0.3$, and 
$\Omega_{\lambda}=0.7$ throughout this paper);
if this light is due to stars, these objects each would have the luminosity
of a substantial galaxy.
There is also a close grouping of faint objects to the north of
the quasar, seemingly embedded in a general extended low-surface-brightness
haze, which also extends somewhat around to the west, culminating in a
narrower finger to the south.  The linear feature is also present in
the Keck LRIS $z$ image, although it is of course not as clearly defined
because of the lower resolution.  However, the diffuse material to the north
is quite clearly seen; the south-pointing ``finger'' $b$ appears here to 
extend directly towards the quasar, whereas, in the WFPC2 image, it seems to 
project back to the west side of the quasar.  The NIRC $K'$ image
agrees with the $z$ image in this respect.  It also shows a faint
apparent extension ($c$) to the north-northwest of the linear feature.  On the
other hand, the linear feature itself near the quasar, so clearly defined in the
WFPC2 image, is not present at all in the $K'$ image, although the bright
knot $d$ $\sim1\farcs5$ west of the quasar lies right in the middle of it.
The diffuse material north of the quasar is still present at $K'$, and
four of the faint objects projected on it ($e$, $f$, $g$, and $h$) now dominate 
the others.

The extended material looks quite different in [\ion{O}{2}] emission,
although there are some connections with features seen on the 
continuum-dominated images.  The strong extension to the south is aligned 
with the ``finger'' $b$ seen in the $z$ and $K'$ images.  The strongest 
off-nuclear [\ion{O}{2}] emission occurs in the same region as the strong 
diffuse continuum emission seen southwest of the quasar in the WFPC2 image, 
which is also close to the tangent point of the linear feature.
Faint diffuse emission is present to the north and northwest, some of which
may be associated with discrete features seen in the continuum images.

\begin{figure}[!bt]
\epsscale{1.0}
\plottwo{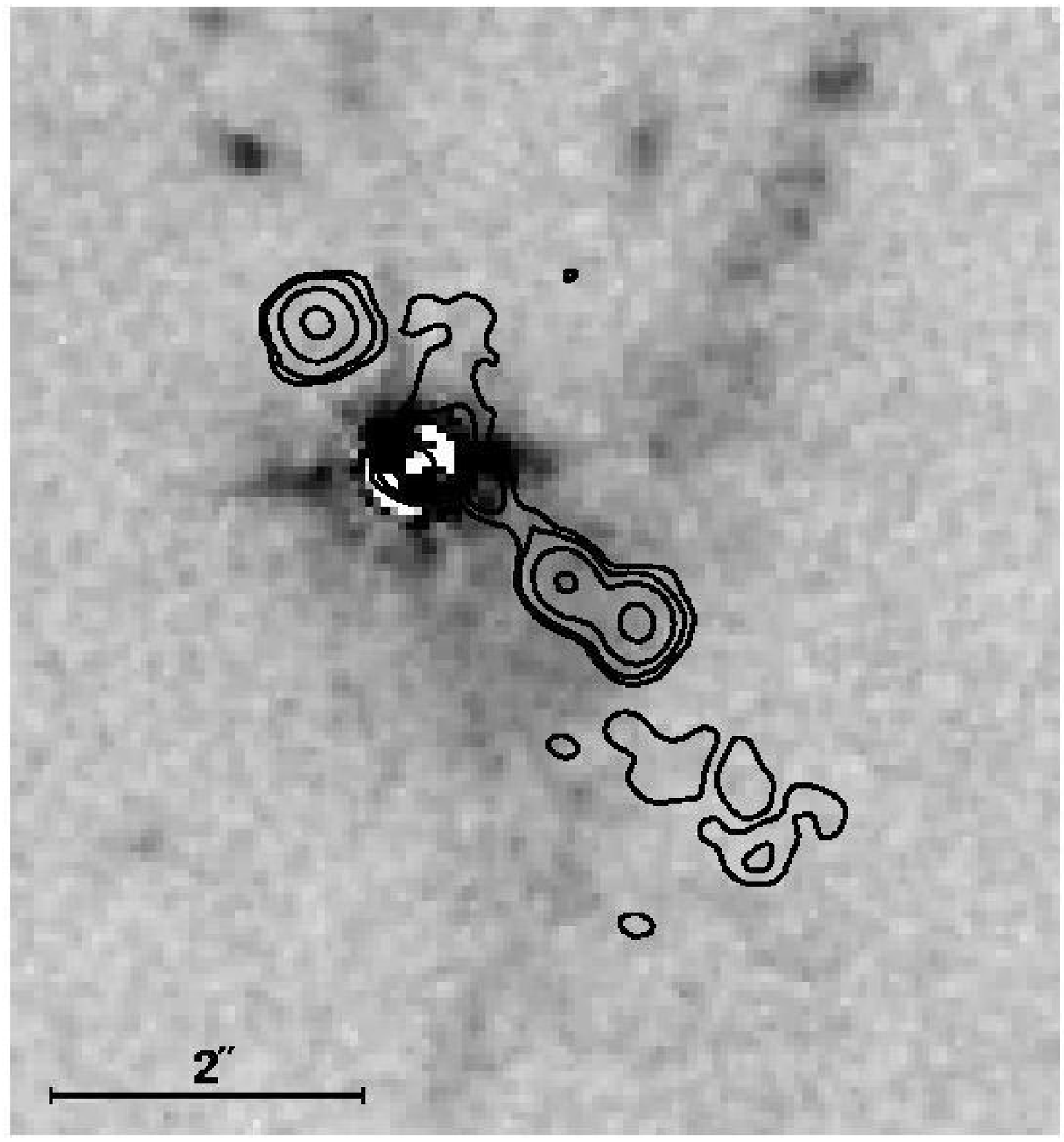}{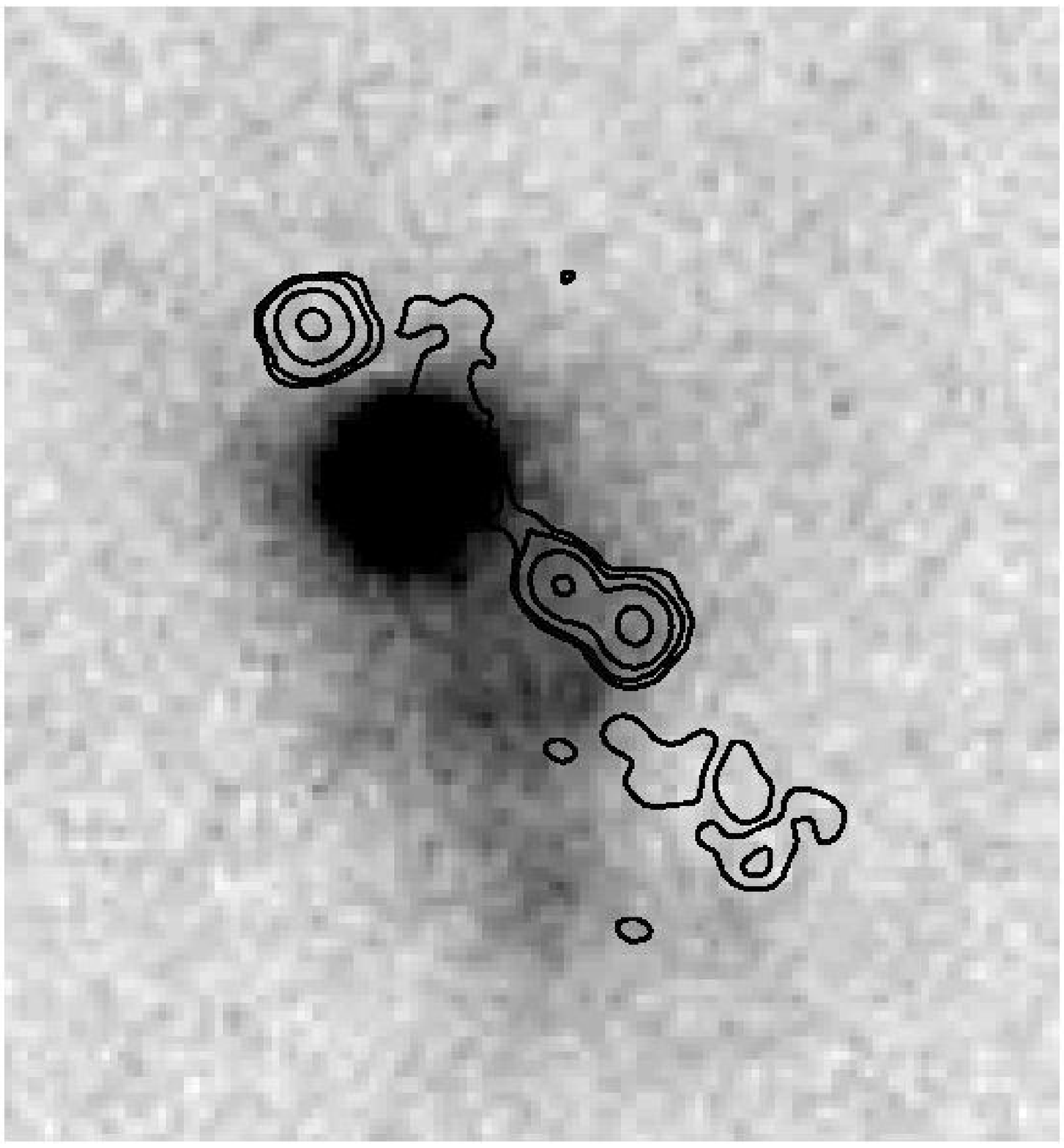}
\caption{The 1662 MHz radio map \citep{spe91} has been superposed on
the {\it HST} WFPC2 F702W image (left panel) and the {\it CFHT} SIS [\ion{O}{2}]
image (right panel).  The radio contours are at 6, 24, 96,
and 240 mJy per beam.}\label{radio}
\end{figure}

3C\,190 is often regarded as a compact steep-spectrum (CSS) source 
\citep[\eg][and references therein]{dev99}; however, the radio morphology
is that of well defined double source with a projected separation of
2\farcs6 \citep{spe91}, corresponding to $\sim20$ kpc, 
so 3C\,190 
can be thought of as a small classical double source, especially considering
likely projection effects.  The radio contours are superposed on the
WFPC2 and [\ion{O}{2}] images in Fig.~\ref{radio}.  While the southwest 
hotspot and associated
radio jet fall in a region of diffuse emission near the point of closest
approach of the linear feature $a$ to the quasar, there is no clear 
correlation between the radio and optical structure on the WFPC2 image.
Some of the strong [\ion{O}{2}] emission to the south of 3C\,190 appears
to fall along the south-eastern side of the radio jet, but the
correlation is not so strong that a physical association between the two
is compelling, at least from the imaging data alone.
%There is a better correlation with some features in the [\ion{O}{2}] image,
%although most of the strong [\ion{O}{2}] emission falls to the east of
%the radio jet.

The evidence from the images indicates that we may be seeing merging
processes on at least two levels going on simultaneously.  The luminous
extension to the south, seen best in the $z$ and $K'$ images, has the
appearance of a major tidal tail, and the extension to the north-northeast
may be a second such feature.  In fact, the geometry of these possible
tails, as seen on the $K'$ image, bears a strong resemblence to the local
merger NGC\,7252 \citep{sch82}.  However, in addition to this evidence for
a major merger, the diffuse luminous material and many small galaxies to
the north also appear to be poised to merge with the host galaxy of 3C\,190.

There remains the 50-kpc-long linear feature seen in the WFPC2 and $z$ images,
which comprises clumps having luminosities and dimensions similar to those
of the many small galaxies in the vicinity.  Is it truly a linear object, 
similar to the so-called ``chain galaxies'' \citep[\eg][]{cow95}
found to be fairly common at
high redshifts?  Or could it be a large edge-on disk galaxy, where the peak
seen in the $K'$ image just to the west of the quasar is the nucleus, which
is hidden by dust at shorter wavelengths?  We turn now to the spectroscopic
observations, from which we can hope to investigate both
the nature and the kinematics of the emitting material.

\subsection{Spectroscopy}
\begin{figure}[!tb]
\epsscale{0.42}
\plotone{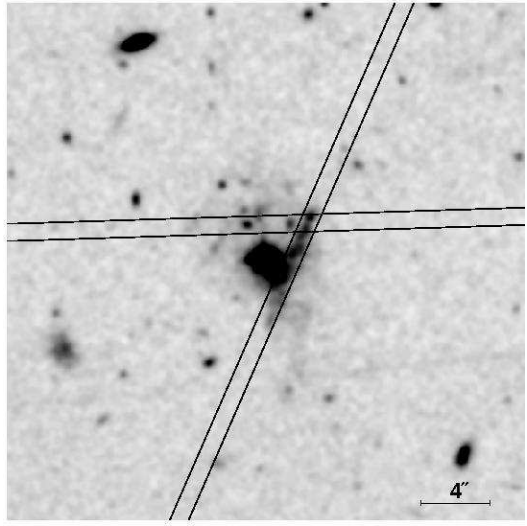}
\caption{Slit positions for the LRIS spectroscopic observations, superposed
on a smoothed version of the WFPC2 image.  Slit positions centered on
the quasar itself are omitted for clarity.}\label{3c190sl}
\end{figure}

The slit positions for all of the spectra except those of the quasar 
itself are shown in Fig.~\ref{3c190sl}.

\subsubsection{The quasar spectrum}
\begin{figure}[!tb]
\epsscale{0.8}
\plotone{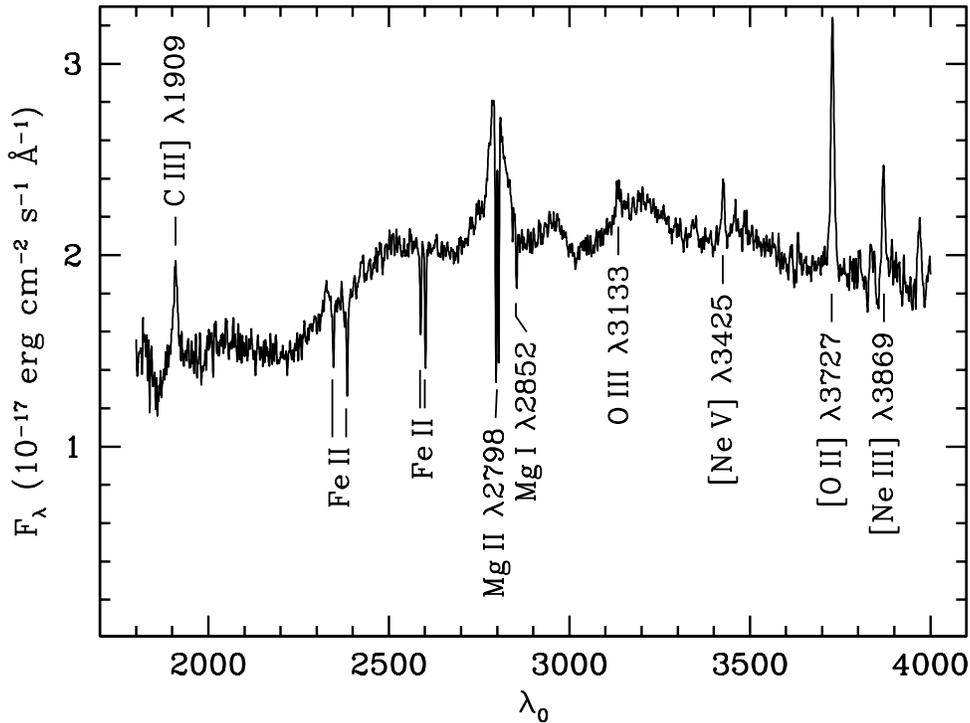}
\caption{The spectrum of 3C\,190 (the quasar itself).  The absorption lines
all appear to come from a single system at close to the quasar redshift.}
\label{qspec}
\end{figure}

The spectrum of the quasar is shown in Fig.~\ref{qspec}.
The decline in the continuum towards the UV is apparent, consistent with
the classification of 3C\,190 as a ``red'' quasar \citep{smi80}.  There are 
also a number of newly detected absorption lines in the spectrum, including 
a very strong \ion{Mg}{2} pair at close to the quasar 
rest frame, which clearly may be produced by gas associated both with 
the dust that is responsible for reddening the quasar and with some of the
extended structure we see in our images.  From narrow [\ion{Ne}{3}]
$\lambda3869$, [\ion{O}{2}] $\lambda\lambda3726$,3729, and \ion{C}{3}]
$\lambda1909$, we obtain a quasar redshift of $1.1946\pm0.0005$.  The
strong \ion{Mg}{2} absorption has a redshift of $1.19565\pm0.00004$,
so it has a radial velocity of $145\pm70$ km s$^{-1}$ in the quasar frame.
Other prominent lines in this same system include \ion{Mg}{1}
$\lambda2852$ and \ion{Fe}{2} $\lambda\lambda2343$,2382,2586,2599.

\subsubsection{The linear feature}

\begin{figure}[!tb]
\epsscale{0.6}
\plotone{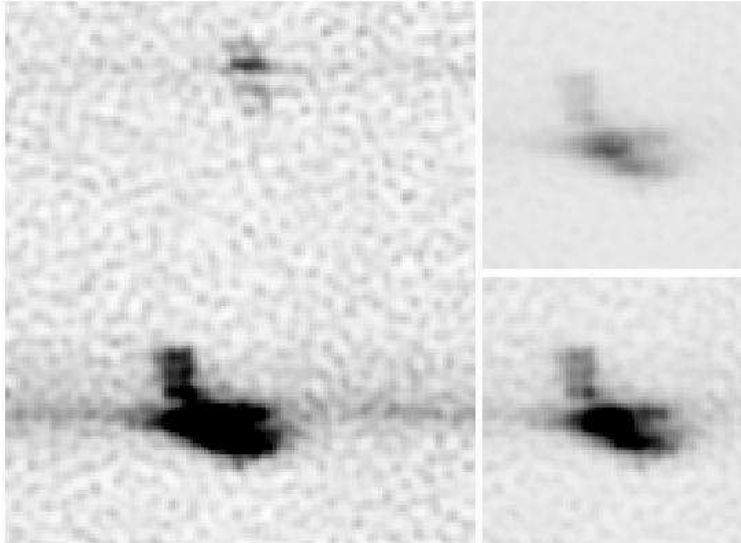}
\caption{Deconvolved spectrum of [\ion{O}{2}] emission in a slit aligned
along the linear feature.  The two panels on the right show lower contrast
versions of the emission near the quasar.  
Note the resolved [\ion{O}{2}] doublet above the gap and the sudden shift 
to higher velocities below the gap, where much broader emission is also seen.
The [\ion{O}{2}] emission at the top of the left-hand panel is from a
faint galaxy $\sim20\arcsec$ north-northwest of 3C\,190.  It has a redshift of 
1.1977, close to that of 3C\,190.  The height of the left-hand panel
corresponds to 30\arcsec, and the north-northwest end of the slit is towards
the top. 
}
\label{decon}
\end{figure}

The region around the [\ion{O}{2}] $\lambda3727$ doublet in our best
two-dimensional spectrum of the linear feature (that of UT 1998 Feb 17)
was deconvolved using a kernel generated from the continuum of the quasar
(for the spatial dimension) and the profile of a strong airglow line
(for the spectral dimension).  After deconvolution, the image was restored
with a Gaussian profile with $\sigma=1$ pixel.  The result is shown in
Fig.~\ref{decon}.
\begin{figure}[!tb]
\epsscale{0.6}
\plotone{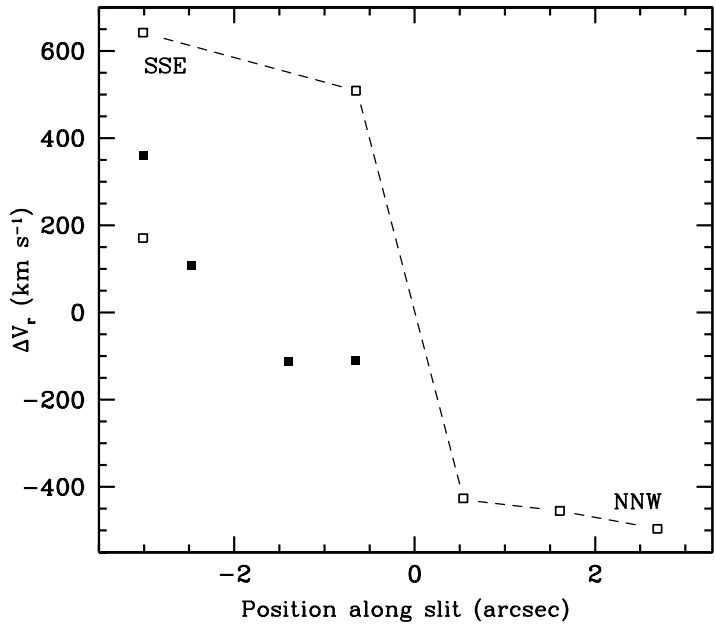}
\caption{Radial velocities in [\ion{O}{2}] emission in a slit aligned
along the linear feature.  The open squares trace emission features
with low internal velocity ($\sigma \sim40$ km s$^{-1}$); the filled squares
trace emission with high internal velocity ($\sigma \sim200$ km s$^{-1}$),
for which the gas has likely has been disturbed by the radio jet.
The dashed lines connect the strongest low-$\sigma$ components.  The spatial
zero point is component $d$, seen in the $K'$ image (see Fig.~\ref{3c190mos},
which corresponds with a gap in the linear feature in the optical.  The
velocity zero point was arbitrarily chosen to be $z=1.194$.  The
north-northwest and south-southeast ends of the velocity curve are indicated.}
\label{oiivel}
\end{figure}

Figure \ref{oiivel} shows velocities derived from this spectrum.  The line
profiles divide into two groups:  those with internal velocities with
$\sigma \sim40$ km s$^{-1}$, which likely are due either to star formation
or photoionization by the quasar of quiescent gas, and those with $\sigma 
\sim200$ km s$^{-1}$, which probably indicates gas that has been shocked 
in some way by interaction with the radio jet plasma.  The 
low-velocity-dispersion gas
shows an abrupt velocity dislocation at the position of the main gap in
the linear feature, which corresponds to component $d$ in the $K'$ image.
The overall impression (as traced by the dashed lines in Fig.~\ref{oiivel})
is that of a disk rotation curve, but with a velocity amplitude twice that
of typical giant disk galaxies in the local Universe.  Alternatively,
the velocities could represent bipolar outflow (or infall), or they could
simply mean that the two parts of the so-called linear feature are distinct
and separate objects.

\begin{figure}[p]
\epsscale{0.9}
\plotone{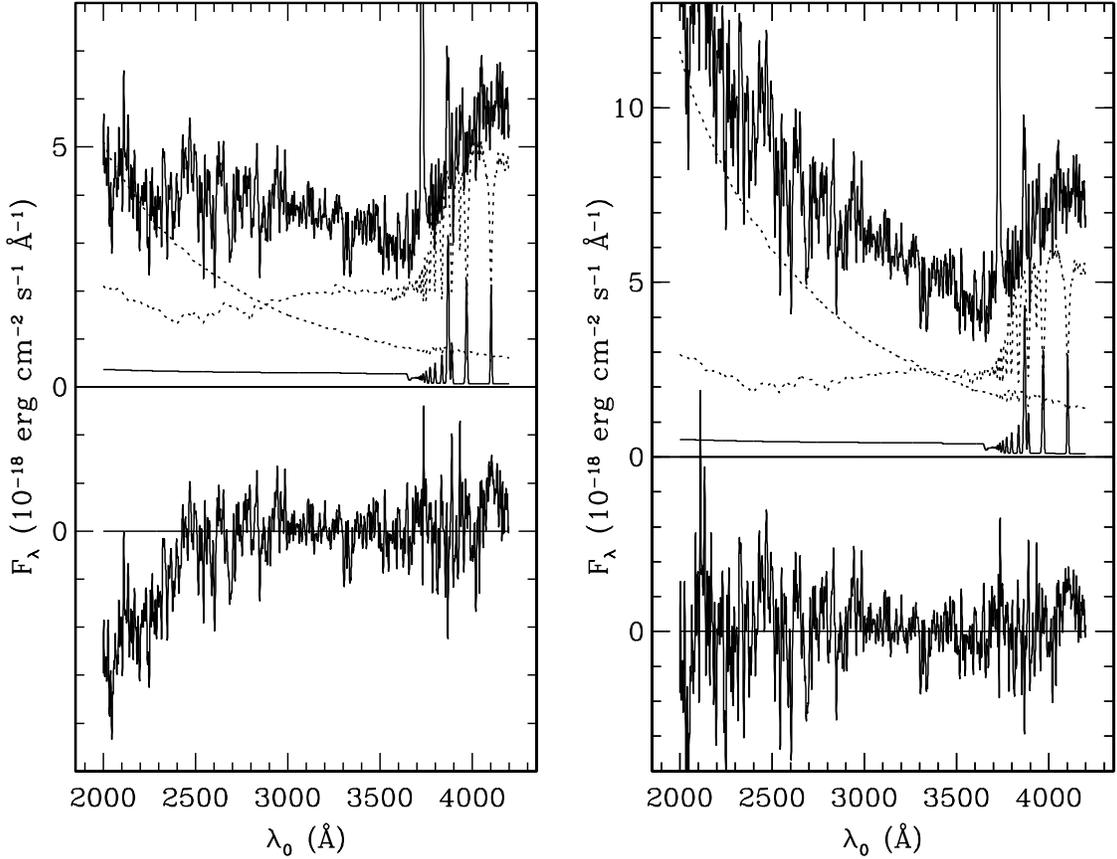}
\caption{Analysis of the spectrum of the linear feature.  In the top-left panel,
the observed summed spectrum of the northwestern portion of the
linear feature is shown as the solid line at the top.  The solid line just
above the zero point is a model for the Balmer and forbidden line emission,
based on assuming an [\ion{O}{2}]/H$\beta$ intensity ratio of 4.0 and
Case B recombination, with an electron temperature of $1.5\times10^4$ K;
further details are given in the text (the simulated profile of the
[\ion{O}{2}] doublet is omitted for clarity, but it is included in the
subtraction mentioned below).  The two dotted spectra are instantaneous 
burst spectral synthesis models \citep{bru96} with ages of
$2\times10^6$ and $4.5\times10^8$ years, scaled so that
their sum matches as well as possible, over the range 2500--4200 \AA, the
observed spectrum after the emission component has
been subtracted.  The residual after subtracting all of these components
from the observed spectrum is shown as the solid
trace in the bottom-left panel, where we have also included a zero-point
line.  Note the steep fall off shortward of
2500 \AA.  The right top and bottom panels are similar, except the observed 
spectrum has
been corrected by dividing by an assumed extinction curve (see text for 
details).  In the top-right panel, the spectral synthesis models have ages of 
$2\times10^6$ and $3.6\times10^8$ years.
}
\label{linspec}
\end{figure}

The summed spectrum of the linear feature to the northwest of the gap
is shown as the solid trace at the top of the left panel in 
Fig.~\ref{linspec}.  There are at least two 
remarkable features of this spectrum:  the inflection in the slope of the 
continuum near the position of the [\ion{O}{2}] $\lambda3727$ doublet, and 
the absence of any evidence for H$\delta$ emission.  We consider this second 
point first.

Although the only emission obvious in this spectrum besides the [\ion{O}{2}]
doublet is the [\ion{Ne}{3}] $\lambda3869$ line, the ratio of [\ion{Ne}{3}]
to [\ion{O}{2}] intensities (0.14) is quite consistent with that found in 
moderately high ionization \ion{H}{2} regions; \eg\ the giant \ion{H}{2} region
in the LMC, 30 Doradus, has a ratio of 0.16 \citep{ken00}.  
If we take 30 Doradus and two other \ion{H}{2} regions listed by \citet{ken00}
as having [\ion{Ne}{3}]/[\ion{O}{2}] ratios similar to that we find for
3C\,190, the average [\ion{O}{2}]/H$\beta$ intensity ratio is $1.82\pm0.47$.
On the other hand, some
\ion{H}{2} regions in the sample of \citet{ken00} have [\ion{O}{2}]/H$\beta$ 
intensity ratios ranging up to $\sim4$ (and even slightly beyond), so we use 
this value to calculate a rough lower limit to the expected H$\delta$ 
intensity.  Taking this lower limit as the actual intensity, and assuming 
Case B recombination
for a temperature of $1.5\times10^4$ K, we can construct a template
emission-line spectrum consisting of the Balmer lines, the
[\ion{O}{2}] $\lambda\lambda3726$,3729 doublet, and the [\ion{Ne}{3}]
$\lambda\lambda3869$,3967 lines (where we simply fit Gaussian profiles
to the forbidden lines and use the fact that the ratio of 
the [\ion{Ne}{3}] $\lambda3967$ photon 
flux to that of the $\lambda3869$ line is set by atomic parameters to
be 1:3).  We subtract this template from the original spectrum.

At this point we could attempt to fit this residual with some combination of
scattered quasar radiation and stellar spectral synthesis models.
Unfortuately, the broad \ion{Mg}{2} line in the quasar spectrum does not
prove to be a useful constraint on the amount of scattered light because of
its relatively low equivalent width with respect to the S/N in the
spectrum of the linear feature.  A better constraint can be obtained by
comparing the flux density from knots in the linear structure with that
of direct radiation from the quasar.  Assuming the depth along the line-of-sight
is comparable to the characteristic diameter in the plane of the sky, and
assuming 100\% scattering efficiency of the incident quasar light, we find the 
contribution of scattered quasar light to be $<10$\% of the total observed
continuum.  As the actual scattering
efficiency is likely to be more like 20\% at best, and projection effects
may be important, we are almost certainly safe in ignoring this scattered
contribution.

Assuming, then, that the residual continuum is dominated by stellar
radiation, we obtain a fair fit over most of the range with
a solar-metallicity stellar population with an age of $\sim2\times10^8$
years, but there remains a steep dropoff shortward of 2500~\AA.
We can do considerably better by using a two-component stellar population,
as shown in the left panel of Fig.~\ref{linspec}, but, if we optimize the
fit over the 2500--4200 \AA\ region, the residual still has too little
flux below 2500 \AA.  This shortage is at least partly caused by an apparent 
dip between 2100~\AA\ and 2400~\AA, which can be seen at some level in the 
original data.  This dip is very likely the 2175~\AA\
dust extinction feature found in the Galactic extinction law, although
\citet{cal94} found it to be generally absent in a sample of 39 starburst
galaxies.  We model the extinction by taking the Galactic extinction law,
modified by including isotropic scattering \citep[see][Fig.~10]{cal94},
and then normalizing this curve to the \citet{cal94} effective extinction
curve for wavelengths far from the 2175~\AA\ feature.  We then adjust the
optical depth to achieve the best cancellation of the feature
(in the spectrum as originally observed), as judged
by eye.  The resulting corrected spectrum is shown as the upper trace in 
the right panel of
Fig.~\ref{linspec}.  We now repeat the fitting and subtraction of emission
lines and nebular thermal continuum, 
as described above.  We fit the remaining continuum as the sum of two
spectral synthesis models:  one with an age of 2 Myr, and
one with an age of 360 Myr.  The residual from subtracting the nebular
component and the two stellar components from the extinction-corrected
spectrum is shown in the bottom right panel of Fig.~\ref{linspec}.

The many uncertainties of this exercise, particularly in estimating both the
wavelength dependence and the amplitude of the extinction correction, prevent
us from claiming any quantitative validity for the results.  However, we
believe that certain qualitative results are fairly robust:  (1.) The
inflection in the observed continuum indicates the presence of a 
few-hundred-Myr-old
population, whose Balmer absorption lines are partially masked by emission
lines.  The inflection is present in the data before the extinction
correction; while the age of this component is dependent on the amount of
extinction and the modeling of any younger population, it almost certain
has to be in the range 2--$5\times10^8$ years.  (2) There is a significant 
level of extinction.  This conclusion is based on the difficulty
of obtaining a satisfactory fit to the observed spectrum using spectral
synthesis models and the evidence for a dip near 2200 \AA.  (3) A very
young stellar component is present, consistent with current ongoing
star formation.  This component is required to obtain a decent fit to the 
spectrum, particularly if any
significant extinction is included, and it would provide the most
logical source of ionization for the observed emission-line spectrum.
Nevertheless, (4) this actively starforming component comprises only a 
small fraction of the stellar mass (the formal value for our specific model 
is $\sim2.5$\%).  If our model is close to being accurate, we estimate a total 
stellar mass of $\sim10^{10} M_{\sun}$ for each of the three brightest regions.

\subsubsection{Other galaxies associated with 3C\,190}

\begin{figure}[tb]
\epsscale{0.6}
\plotone{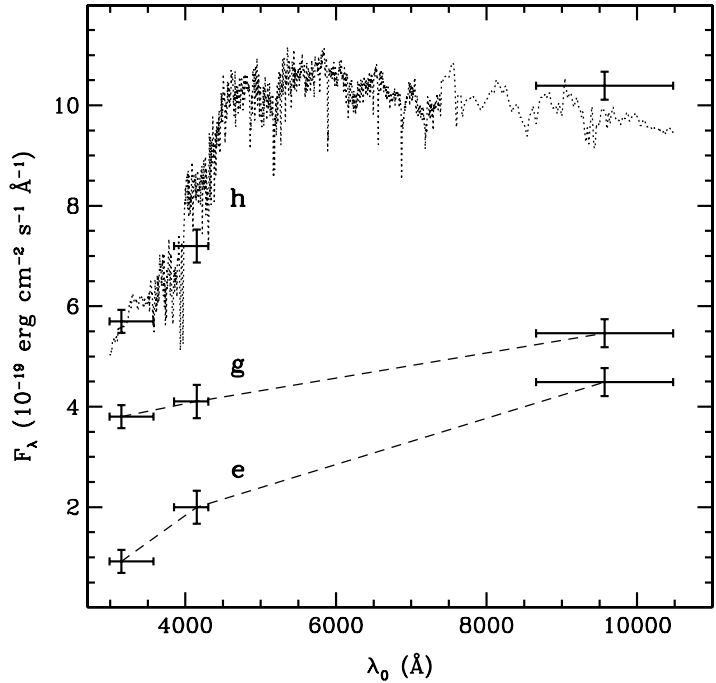}
\caption{Spectral-energy distributions of three galaxies within 6\arcsec\
of 3C\,190.  The vertical bars show 1 $\sigma$ photometric uncertainties
based on sky variations; these do not include systematic uncertainties from
adjacent objects, aperture corrections, and photometric zero points.  The
horizontal bars indicate the filter half-transmission points.
The values for galaxies $g$ and $h$ have been displaced upwards by 2 and
4 units, respectively.
A Bruzual \& Charlot (1996) spectral-synthesis model with an age of 4 Gyr 
has been placed on the photometry for galaxy $h$ for comparison.
}
\label{galsed}
\end{figure}

There are a number of galaxies clustered around 3C\,190, mostly to
the north.  The brightest of these have $M_B\sim-20$.  Three galaxies
very close to 3C\,190 are detected and well-resolved on all three of
our broad-band images, and their SEDs are shown in Fig.~\ref{galsed}.
Of these, galaxy $h$ comes close to fitting an essentially unreddened
old stellar population at the quasar redshift, whereas galaxies $e$ 
and $g$ would be consistent with reddened younger populations.

One of our slit positions from 1998 February 15 covered two of the galaxies
north of 3C\,190 as well as the northern part of the linear feature (see
Fig.~\ref{3c190sl}).  The eastern of these is seen on the {\it HST} image
to be part of a close triplet, which is included within the diffuse object
labeled $f$ in the Keck NIRC image (Fig.~\ref{3c190mos}).  This object
shows [\ion{O}{2}] $\lambda3727$ emission at $z=1.1942$, on an extremely
weak continuum.  The other galaxy was $e$ in Fig.~\ref{3c190mos} (see also
Fig.~\ref{galsed}); it has
a stronger continuum, with a slight slope upwards towards the red across
the rest-frame UV range,
%, with an almost constant flux density of $\sim10^{-18}$
%erg cm$^{-2}$ s$^{-1}$ \AA$^{-1}$ in the rest-frame UV 
and it shows [\ion{O}{2}] $\lambda3727$ and
[\ion{Ne}{3}] $\lambda3869$ emission at $z=1.1935$.  The faint emission-line
galaxy $\sim20$\arcsec\ to the northwest and fortuitously on our slit
placed through the linear feature (see Fig.~\ref{decon}) has $z=1.1983$.
(It also has very high ionization, with [\ion{Ne}{5}] $\lambda3426$ as
strong as [\ion{Ne}{3}] $\lambda3869$.)  

These observations indicate that
we can safely assume that most of the dozen or so small galaxies within 
$\sim6$\arcsec\ ($\sim50$ kpc) radius of 3C\,190 are associated, and it is 
likely that many at larger distances are as well.  The galaxies in the
immediate vicinity of the quasar host form a compact group which must
almost certainly shortly merge with it. 

We can make an estimate of the total stellar mass in all of
the objects in the immediate vicinity of 3C\,190 from the PSF-subtracted
$K'$ image.  We use a 4\arcsec-radius aperture, centered 0\farcs8 east and
1\farcs6 north of the quasar, which includes the quasar host galaxy as well
as objects $d$--$h$.  If we assume that there is no significant extinction 
at this wavelength (restframe 9600 \AA) and
that the stellar population can be modelled as a single burst 5 Gyr old
with a Salpeter initial mass function, we will have a fairly firm lower limit
to the stellar mass.  By scaling a \citet{bru96} model to the observed
flux density, we obtain a mass of $\sim2\times10^{11} M_{\sun}$.  The mass 
obtained is
not too sensitive to the age of the stellar population for ages $\ge1$ Gyr;
\eg\ if we assume an age of 1 Gyr, we only double the mass.  Very little of
the mass can be in stars much younger than 1 Gyr unless the extinction
in regions containing such stars is extremely high.  If our quasar PSF 
subtraction is reasonably accurate, the fraction
of the stellar mass within 1\arcsec\ (= 8 kpc) of the quasar (corresponding
roughly to the quasar host galaxy) is $\sim0.25$.

\section{Discussion}
\subsection{Summary and Interpretation of the Observations}

There is a bewildering variety of phenomena found in the vicinity of 
3C\,190, and the manner in which each relates to the rest is not at all
obvious.  We summarize here the main features and what we take to be their most
plausible interpretations.  

The narrow, low-surface-brightness features $b$ and $c$ are seen
best in the Keck $K'$ image but are also visible in the $z$ image.
If these are tidal tails, as seems possible, they indicate a past major merger.

The tight grouping of small galaxies, mostly on the north side, is
projected on low-surface-brightness extended material (seen best on the
{\it HST} and $z$ images).  Our spectroscopy shows that at least two of
these are associated with 3C\,190, and the high local density of these
objects suggests that most of the others are associated as well.  The
richness and compactness of this group indicates that the 3C\,190 host
galaxy is subjected to a continuing rain of minor (and some fairly major)
mergers.

The high surface brightness, high-velocity-dispersion emission-line gas
extends mostly to the south. It appears to start out in almost the same 
direction as the low-surface-brightness feature $b$, but it is probably 
associated instead in some way with the radio emission, though it is only 
loosely correlated with it in position.

The long linear feature $a$ remains enigmatic, even though we have learned 
quite a bit about its stellar content and velocity structure. 
It is still well defined in our $z$ image,
which is dominated by light longward of 4000 \AA\ in the rest frame.  This means
that the stars a few $\times10^8$ years old that we find in our spectroscopy
of the feature are dynamically associated with it, rather than being part
of the general diffuse background seen around the quasar.  At shorter
wavelengths, the feature is dominated by the large star-forming complexes that
provide the ionization for the low-velocity-dispersion (LVD) gas.  
It is the apparent velocity structure for this LVD gas that makes the
interpretation of this feature difficult.  The velocity curve is consistent 
with that of a disk embedded in an extremely massive halo ($\sim10^{12} 
M_{\sun}$ within a 25 kpc radius); but the combination of the extreme
rarity of such objects, the association with (but clear offset from) 3C\,190,
and the necessity that we would be observing the object almost exactly
edge on, makes this interpretation difficult to accept.  Collimated outflow
makes even less sense, given the clear indication of a significant mass
of stars and the fact that the supposed outflow would have to
originate, not at the quasar, but at or near object $d$.  On the whole,
a close connection with the ``chain galaxies'' seen frequently at high
redshift \citep{cow95} seems most likely, although these are themselves
poorly understood.  Furthermore, this interpretation does nothing to explain the
observed velocity structure.  If the feature is interpreted as two separate
chain galaxies at different velocities, then their alignment
with each other would be fortuitous.  A remaining, but remote, possibility is 
that the
LVD [\ion{O}{2}] emission at the northwest and southeast ends of the feature
have independent physical origins, the former being the result of young
stars, the latter perhaps being due to precursor ionization from shocks 
associated with the radio jet.

\subsection{The Formation of Ellipticals}

A prediction of the standard cold dark matter (CDM) scenario for galaxy 
formation is that small baryonic structures form first and then merge to form 
larger structures.  But a persistent problem of the simplest ``building-block'' 
models, when applied to the formation of spheroids, is the observed
correlation between luminosity and color in present-day ellipticals and bulges,
implying a correlation between mass and metallicity 
\citep[e.g.,][]{bow92,ell96}.  
If, for example, most of the stars now in cluster ellipticals were already 
formed in dwarf-galaxy-sized systems, how would the stars acquire the 
metallicities characteristic of the mass of the galaxy into which they would 
eventually be incorporated?  In fact, if the ``building blocks'' are both 
small and isolated at the time they form most of their stars, they will not 
easily retain the enriched gas needed to form the later 
generations of higher metallicity stars.

What we can see happening around 3C\,190 is indeed the formation of a large
number of galaxies with a range of sizes.  Some, such as the emission-line
galaxy fortuitiously found on our slit some 20\arcsec\ from 3C\,190 may
simply indicate that 3C\,190 is in a rich environment, possibly a cluster
in formation.  Others, closer in, appear to be in the process of merging with 
the 3C\,190 host galaxy.  But these latter are not isolated entities:  some 
objects are incorporated into a coherent structure with a characteristic
dimension of $\sim50$ kpc (whether a disk, chain galaxy, or something else);
others appear to be distributed more randomly, but they are still found within
a common luminous envelope.  This means that the star formation in these
individual building blocks is likely taking place within an already established
potential well sufficiently deep to retain the enriched gas for successive
generations of star formation.  Furthermore, we have observational evidence
within the star-forming regions of the linear feature for at least one major
previous generation of stars.

All of this means that the distinction between a monolithic picture of
bulge or elliptical-galaxy formation {\it \`{a} la} \citet{egg62} and the
bottom-up CDM view may not be as clear cut 
as is sometimes supposed.  In the objects within the immediate environment of 
3C\,190, processes of star formation and assembly into larger aggregates 
appear to overlap in time and to influence each other, leading to
complex feedback processes during the formation of what almost certainly
will become an elliptical or cD galaxy at the present epoch.

\acknowledgments

We thank Josh Barnes, Len Cowie, and John Kormendy for a helpful discussions,
and we thank Joel Aycock and Barbara Shaefer for carrying out the
$z$-band imaging as a Keck service observing program.
Support for this work was provided by NASA through Grant No.\ GO-06491.01-A
from the Space Telescope Science Institute, which is operated by AURA, Inc.,
under NASA Contract No.\ NAS 5-26555.  Additional support was provided by
NSF under grant AST95-29078.
This research has made use of the NASA/IPAC Extragalactic Database (NED) 
which is operated by the Jet Propulsion Laboratory, California Institute of 
Technology, under contract with the National Aeronautics and Space 
Administration.

\clearpage


\begin{thebibliography}{}
\bibitem[Bower, Lucey, \& Ellis(1992)]{bow92} Bower, R. G., Lucey, J. R., \&
Ellis, R. S. 1992, \mnras, 254, 601
\bibitem[Bruzual \& Charlot(1996)]{bru96} Bruzual A., G. \& Charlot, S. 1996,
   unpublished \linebreak[0] [ftp://gemini.tuc.noao.edu/pub/charlot/bc96]
\bibitem[Calzetti et al.(1994)]{cal94} Calzetti, D., Kinney, A. L., \& 
Storchi-Bergmann, T. 1994, \apj, 429, 582
\bibitem[Casali \& Hawarden(1992)]{cas92} Casali, M. M., \& Hawarden, T. G. 
1992, UKIRT Newsletter, 4, 33
\bibitem[de Vries et al.(1999)]{dev99} de Vries, W. H., O'Dea, C. P., Baum,
S. A., \& Barthel, P. D. 1999, \apj, 526, 27
\bibitem[Chambers et al.(1987)]{cha87} Chambers, K. C., Miley, G. K., \&
van Breugel, W. 1987, \nat, 329, 604
\bibitem[Chambers \& Miley(1990)]{cha90} Chambers, K. C., \& Miley, G. K. 1990,
The Evolution of the Universe of Galaxies, ed.~R. G. Kron, (San Francisco:
Astronomical Society of the Pacific), p.~373
\bibitem[Cowie et al.(1995)]{cow95} Cowie, L. L., Hu, E. M., \& Songaila, A.
1995, \aj, 110, 1576
\bibitem[Eggen, Lynden-Bell, \& Sandage(1962)]{egg62} Eggen, O. J.,
Lynden-Bell, D., \& Sandage, A. R. 1962, \apj, 136, 748
\bibitem[Ellis et al.(1996)]{ell96} Ellis, R. S., Colless, M., Broadhurst, T., 
Heyl, J., \& Glazebrook, K. 1996, \mnras, 280, 235
\bibitem[Ferrarese \& Merritt(2000)]{fer00} Ferrarese, L., \& Merritt, D. 2000, 
\apj, 539, L9
\bibitem[Gebhardt et al.(2000)]{geb00} Gebhardt, K., et al.\ 2000, \apj, 
539, L13
\bibitem[Kennicutt et al.(2000)]{ken00} Kennicutt, R. C., Jr., Bresolin, F.,
French, H., \& Martin, P. 2000, \apj, 537, 589
\bibitem[McCarthy et al.(1987)]{mcc87} McCarthy, P. J., van Breugel, W.,
Spinrad, H., \& Djorgovski, S. 1987, \apjl, 321, L29
\bibitem[Massey \& Gronwall(1990)]{mas90} Massey, P., \& Gronwall, C. 1990,
\apj, 358, 344
\bibitem[Matthews \& Soifer(1994)]{mat94} Matthews, K., \& Soifer, B. T. 1994,
in Infrared Astronomy with Arrays:  the Next Generation, ed. I. McLean 
(Dordrecht: Kluwer), p.~239
\bibitem[Oke et al.(1995)]{oke95} Oke, J. B., et al. 1995, \pasp, 107, 375
\bibitem[Ridgway \& Stockton(1997)]{rid97} Ridgway, S. E., \& Stockton, A.
1997, \aj, 114, 511
\bibitem[Schweizer(1982)]{sch82} Schweizer, F. 1982, \apj, 252, 455
\bibitem[Simpson \& Rawlings(2000)]{sim00} Simpson, C., \& Rawlings, S. 2000,
\mnras, in press [astro-ph/0005570]
\bibitem[Smith \& Spinrad(1980)]{smi80} Smith, H. E., \& Spinrad, H. 1980,
\apj, 236, 419
\bibitem[Spencer et al.(1991)]{spe91} Spencer, R. E., et al.\ 1991, \mnras,
250, 225
\bibitem[Stockton(1999)]{sto99} Stockton, A. 1999, in Galaxy Interactions at
   Low and High Redshift, IAU Symp. 186, eds. D. Sanders \& J. Barnes
   (Dordrecht: Kluwer), p.~311
\end{thebibliography}
\end{document}